\documentclass[cits]{PoS}

\title{Submm-bright QSOs at $z\sim2$: signposts of co-evolution at high $z$}

\ShortTitle{Submm-bright QSOs at $z\sim2$: signposts of co-evolution at high $z$}

\author{\speaker{Francisco J. Carrera}, Anuar K. Al\'\i\\
        Instituto de F\'\i{}sica de Cantabria (CSIC - U. de Cantabria), Spain\\
        E-mail: \email{carreraf@ifca.unican.es},\email{anuarkhan@ifca.unican.es}}

\author{Mathew J. Page, Myrto Symeonidis\\
MSSL-UCL, United Kingdom\\
E-mail: \email{mjp@mssl.ucl.ac.uk},\email{msy@mssl.ucl.ac.uk}}
\author{Jason Stevens, Jos\'e Manuel Cao Orjales\\
Centre for Astrophysics Research, U. Hertfordshire, United Kingdom\\
E-mail: \email{J.A.Stevens@herts.ac.uk}, \email{j.cao-orjales@herts.ac.uk}}

\abstract{We have assembled a sample of 5 X-ray and submm-luminous $z\sim2$ QSOs which are therefore both growing their central black holes through accretion and forming stars copiously at a critical epoch. Hence, they are good laboratories to investigate the co-evolution of star formation and AGN. We have performed a preliminary analysis of the AGN and SF contributions to their UV-to-FIR SEDs, fitting them with simple direct (disk), reprocessed (torus) and star formation components. All three are required by the data and hence we confirm that these objects are undergoing strong star formation in their host galaxies at rates 500-2000~$M_\odot/$y. Estimaes of their covering factors are between about 30 and 90\%. In the future, we will assess the dependence of these results on the particular models used for the components and relate their observed properties to the intrinsice of the central engine and the SF material, as well as their relevance for AGN-galaxy coevolution.}

\FullConference{Nuclei of Seyfert galaxies and QSOs - Central engine \& conditions of star formation\\
		November 6-8, 2012\\
		Max-Planck-Insitut f\"ur Radioastronomie (MPIfR), Bonn, Germany}

\begin{document}

\begin{figure}
\center{\includegraphics[width=0.6\textwidth]{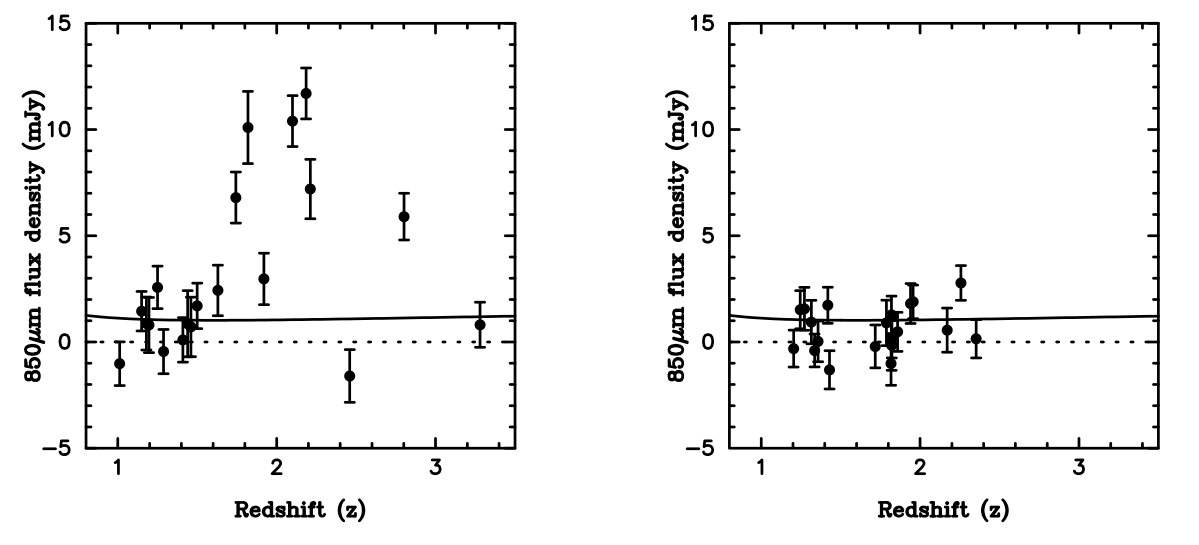}}
\caption{SCUBA 850~$\mu$m flux versus redshift for the X-ray-absorbed QSO sample (left) and for the control sample of X-ray-unabsorbed QSO (right) showing their very different submm properties}
\label{SubmmAbsUnabs}
\end{figure}

\section{Introduction: our sample}

We initially selected from {\sl Rosat} data \cite{Page2000,Page2001a} a sample of around twenty bona-fide broad line quasars which showed X-ray absorption. They were chosen at the redshifts $z\sim1-3$ and X-ray luminosities $L_X\sim L*$ where the maximum contribution to the X-ray background is expected to come. Such high X-ray luminosities ($L_X\sim10^{44}$~erg/s) imply very strong growth of their central SuperMassive Black Holes (SMBH by accretion). We also selected a similarly-sized control sample of X-ray unabsorbed broad line quasars, which constitute 85\% of the population of AGN at those redshifts and luminosities. 

SCUBA submillimetre (submm) photometry \cite{Page2001b} revealed a dramatic difference between the X-ray absorbed and the X-ray unabsorbed control sample (Fig~\ref{SubmmAbsUnabs}): X-ray-absorbed $z>1.5$ QSOs tended to be very submm bright, while very few (if any) of the control sample QSOs were detected. Using typical QSO SEDs this submm emission is very unlikely to come from the central engine. So, our X-ray absorbed QSOs are both bulding up their SMBH vigorously and forming stars copiously, but only at $z>1.5$ \cite{Page2004,Stevens2005}. Based on the relative percentages of X-ray-absorbed and unabsorbed QSO, this appears to be a transitory phase in the life of a broad-line QSO, encompassing about 15\% of its active phase.

Dropping from our initial sample of submm-bright QSOs one radio loud object (because its submm emission might be contaminated by the jet emission) we were left with our current sample of five QSOs. We obtained broad-band X-ray observations of these objects with {\sl XMM-Newton}, that confirmed the narrow-band measurements of their X-ray emission and showed that the only physically-consistent explanation of their X-ray-absorbed spectra but optically broad-line nature is absorption by intervening ionized material in which the dust would not survive \cite{Page2011}. This is consistent with the presence of ionized absorption in their rest-frame UV-optical spectra. About $\sim4$\% of the radiative power could turn into kinetic luminosity of the outflows, which is about the right amount of feedback required to reproduce the local $M-\sigma$ relation. Again, this is consistent with the hypothesis that these X-ray-absorbed QSOs represent a transition phase between obscured accretion and the luminous QSO phase.

SCUBA maps at 450 and 850~$\mu$m \cite{Stevens2004,Stevens2010} showed that these five QSOs were in the centres of $\sim\times 2-4$ overdensities of submm galaxies (SMG). If these SMGs were at the same redshifts as the central QSOs, they would be each an ULIRG in its own right, forming structures of 100s of kpc. Some show disturbed morphologies indicative of major galaxy mergers. So, these QSOs are really signposts of overdensities in the Universe, where strong star formation is happening.

In one case (RX~J0941) we have photometric redshift evidence for the association of the SMG to the central QSO \cite{Carrera2011}. From fits to their opt-NIR photometry and scaling typical star-forming SEDs to their submm fluxes, we found that they are massive galaxies $\log(M_*/M_\odot)\sim11.5\pm0.2$ with high IR luminosities $L_{IR}\sim10^{13}L_\odot$ in the U/HLIRG range, implying star formation rates SFR$\sim2000 \,M_\odot$/y. The SMGs are not detected in X-rays, setting strong upper limits to their SMBH mass, which is about a factor of six below the local BH-galaxy-mass relation. These galaxies are mature and have very limited scope for additional growth ($M_{gas}\sim5\times10^{10}M_\odot$), with their SMBH lagging the galaxy formation, but with plenty of room for additional buildup: they would reach the local BH-galaxy-mass relation accreting only $\sim 1.5$\% of the gas mass.

\section{New {\sl Herschel} data: the QSO SEDs}

We were awarded {\sl Herschel} AO-1 PACS (100, 150$\mu$m) and SPIRE (250, 350, 500$\mu$m) observations of these five QSOs and the fields around them which, together with existing X-ray-to-submm data, will considerably further our knowledge of these objects, their environments and the relationship between them. We also have {\sl Herschel} data for a related project to investigate the $z<1.5$ X-ray-absorbed QSOs in the original sample, and on-going SCUBA-2 observations of a new {\sl XMM-Newton}-selected sample of X-ray-absorbed QSOs \cite{Streblyanska2013}, to enlarge our sample and quantify better the occurrence of this type of objects among the general population of QSOs.

\begin{figure}
\hbox{
\includegraphics[width=0.5\textwidth]{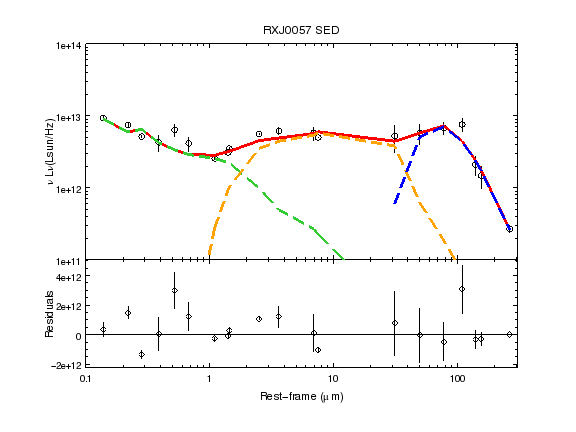}
\includegraphics[width=0.5\textwidth]{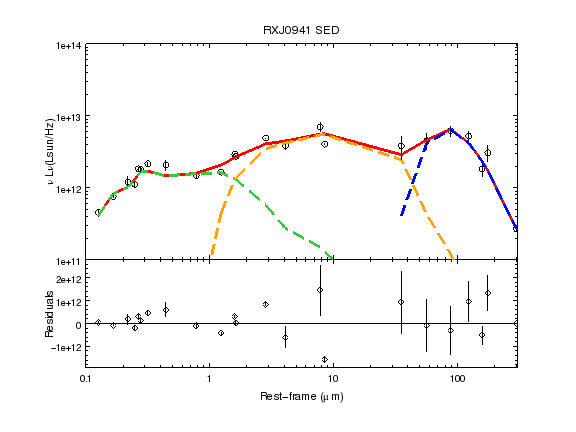}
}
\hbox{
\includegraphics[width=0.5\textwidth]{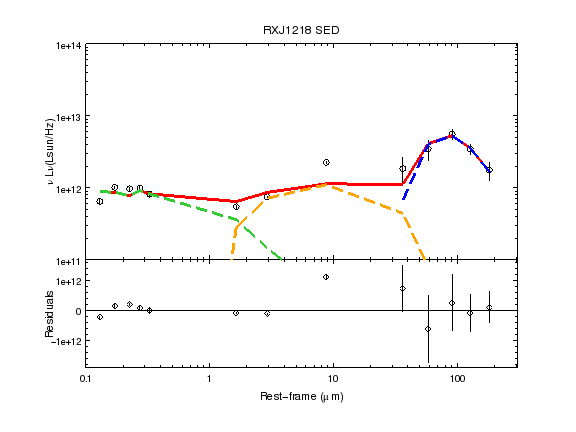}
\includegraphics[width=0.5\textwidth]{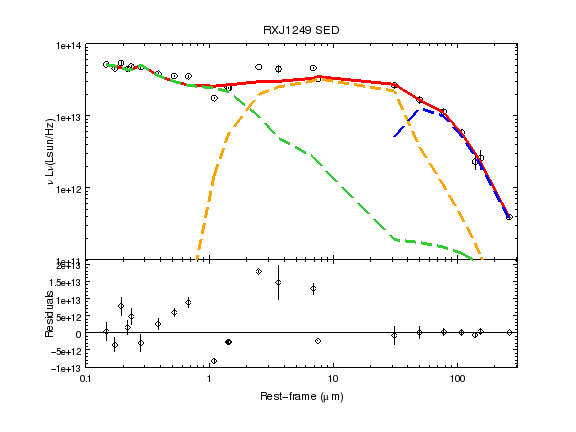}
}
\center{\includegraphics[width=0.5\textwidth]{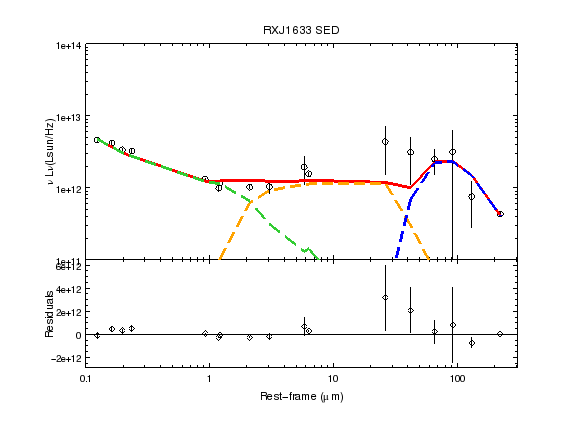}}
\caption{QSO SEDs (open circles), their best fits and residuals in $\nu L_\nu (L_\odot)$ versus rest-frame $\lambda (\mu$m): full model (red solid line), absorbed disk (MRR08, green dashed line), torus (MRR08, orange dashed line) and greybody (blue dashed line)}
\label{FigSEDfits}
\end{figure}

In this presentation we will concentrate on our preliminary analysis of the SEDs of the central QSOs, assembled from our own and archive data (OM, SDSS, 2MASS, WISE), spanning from rest-frame UV to the FIR. The SEDs are shown in Fig.~\ref{FigSEDfits} with a typical number of 17-20 data points, which cover reasonably well the $<10\,\mu$m and $>30\,\mu$m spectral ranges. They are quite flat in the 0.1 to 100~$\mu$m rest-frame range, showing a characteristic decline at longer wavelengths, typical of the thermal origin of the star formation emission expected to dominate at those wavelengths. The emission below $\sim30\,\mu$m is clearly dominant for RX~J1249, which is a very bright QSO. Conversely, for RXJ~J1218 there is a clear thermal-like bump above that wavelength, indicative of a stronger star formation contribution with respect to the AGN emission.

In order to quantify these qualitative impressions, we have fitted the rest-frame 0.1215 to 400~$\mu$m SED with a combination of three components: 1) A direct disk component ({\tt newagn4} in \cite{MRR08}) absorbed by nuclear intervening material. This introduces two free parameters: $A_V$ and the integrated 0.5-250~$\mu$m luminosity $L_{disk}$; 2) A reprocessed torus component ({\tt dusttor} in \cite{MRR08}), with a single free parameter: the integrated 1-300~$\mu$m luminosity $L_{tor}$; 3) A star formation component, parameterized with a greybody model, with three free parameters: $\beta$, the temperature $T$ and the integrated 40-500~$\mu$m luminosity $L_{FIR}$ (which can be directly converted to SFR). Given the high luminosity of this objects and their low obscuration in the rest-frame optical-UV suggest very little contribution from "quiescent" emission from their host galaxies, so we have not introduce an additional component to take this into account.

In total, our models have six free parameters, or about 11-14 degrees of freedom. The best fits are also shown in Fig.~\ref{FigSEDfits}. The fits are reasonable overall, but perhaps some additional "warm dust" component might be needed, for example for RX~J1249. We have done some preliminary fits with other torus models \cite{Nenkova2008} and the corresponding torus luminosities appear to be very similar. We give the best fit parameter values in Table~\ref{TabSEDfits}, estimating the covering factor $CF$ from the ratio between the torus and disk luminosities $CF\sim L_{tor}/L_{disk}$ and the SFR$(M_\odot/$y$)=1.7217\times10^{-10}L_{FIR}$ \cite{Kennicutt1998}.

\begin{table}
\begin{tabular}{llcccrrrr}

Name & $z$ & $L_{disk}/L_\odot$ & $L_{tor}/L_\odot$ & $L_{FIR}/L_\odot$ & $\beta$ & $T/$K & $CF$ & SFR/$M_\odot/$y \\
\hline
RX~J0057  & 2.19 &  $1.84\times10^{13}$  & $1.60\times10^{13}$ & $7.31\times10^{12}$ & 34 & 1.99 &  0.87 & 1300\\
RX~J0941  & 1.82 &  $1.46\times10^{13}$  & $1.37\times10^{13}$ & $6.81\times10^{12}$ & 30 & 1.89  & 0.94 & 1200\\
RX~J1218  & 1.74 &   $7.24\times10^{12}$ & $2.77\times10^{12}$ & $6.21\times10^{12}$ & 36 & 1.02  & 0.38 & 1100\\
RX~J1249  & 2.21 &   $2.42\times10^{14}$ & $8.56\times10^{13}$ & $1.26\times10^{13}$ & 54 & 1.03 &  0.35 & 2200\\
RX~J1633  & 2.80 &   $9.53\times10^{12}$ & $3.23\times10^{12}$ & $2.71\times10^{12}$ & 36 & 1.03  & 0.34  & 500\\
\end{tabular}
\caption{Best fit parameters for the three-component fit to the SEDs of our five QSOs}
\label{TabSEDfits}
\end{table}

\section{Conclusions}

We have assembled a sample of five type 1 QSOs absorbed in X-rays which are growing fast their central SuperMassive Black Holes (SMBH), at the redshifts and luminosities where the maximum contribution to the XRB happens. They are also bright at submillimetric (submm) wavelengths, indicating that they are also forming stars at high rates ($\sim 1000M_\odot/$y.

Their X-ray absorption is from ionized material. There is also evidence for intrinsic absorption in the UV, with strong outflows which, in principle, could provide sufficient feedback to place these objects in the observed $M-\sigma$ relation.

These QSOs are in the centres of overdensities of submm galaxies (SMG). In at least one case (RX~J0941) we have confirmed the association between the central QSO and the SMGs around it. The latter have mature massive galaxies, but their central SMBH (if any) are undergrown with respect to their host galaxies.

We have obtained five-band (PACS and SPIRE) {\sl Herschel} observations of these objects and their environments to complete our picture about them.

In this presentation we show preliminary results on the rest-frame UV-to-FIR SEDs of the central QSOs. We have fitted them with some standard templates for the direct disk emission (extincted by nuclear material), for the re-processed torus emission and for star-formation. We find that all three components are needed for all objects. We confirm that the rest-frame FIR emission is due to strong star formation, with SFR$\sim1000M_\odot$/y (with a range of $500-2500M_\odot/$y). Comparing the re-processed with the direct emission we find covering factors of the order of 50\% (with a range 30-90\%). We are evaluating how much these results depend on the particular model used for the characterization of the different components and whether additional components (such as quiescent emission from the host galaxy or additional "warm dust" emission) are required.  We will next assess the relation of those observational results with intrinsic properties of the QSO, such as its mass or its accretion rate.

Together with the existing and forthcoming X-ray-UV-optical-NIR-MIR-submm information on these fields we will try to understand the nature of these extreme objects and their environments, and assess their relevance for models for co-evolution of SMBH and their host galaxies and the formation of clusters.

\end{document}